\documentclass[12pt]{article}
\usepackage{epsfig}
\usepackage{axodraw}
\begin{document}
\renewcommand{\textfraction}{0.0}
\renewcommand{\topfraction}{1.0}
\renewcommand{\bottomfraction}{1.0}
\newcommand{\aend}{}

%
\catcode`\@=11
\@addtoreset{equation}{section}
\def\theequation{\thesection.\arabic{equation}}
\catcode`\@=12
\newcommand{\be}{ \begin{equation}}
\newcommand{\ee}{ \end{equation}  }
\newcommand{\bea}{ \begin{eqnarray}}
\newcommand{\eea}{ \end{eqnarray}  }
\newcommand{\bi}{\bibitem}
\newcommand{\rd}{ \mbox{\rm d} }
\newcommand{\rD}{ \mbox{\rm D} }
\newcommand{\re}{ \mbox{\rm e} }
\newcommand{\rO}{ \mbox{\rm O} }
\newcommand{\erf}{\mbox{\rm erf}}
\newcommand{\diag}{\mbox{\rm diag}}

\renewcommand{\floatpagefraction}{0.8}

\def\I{\cite{schro}}
\def\del{\partial}
\def\SF{Schr\"odinger functional }
\def\cb{\bar{c}}
\def\q{\tilde{q}}
\def\c{\tilde{c}}

\def\rmd{{\rm d}}
\def\rmD{{\rm D}}
\def\rme{{\rm e}}
\def\rmO{{\rm O}}
\def\tr{{\rm tr}}

 
\def\gms{g_{\ms}}
\def\gmsbar{g_{\msbar}}
\def\gbar{\bar{g}}
\def\gbarms{\gbar_{\ms}}
\def\gbarmom{\gbar_{\rm mom}}
\def\ms{{\rm MS}}
\def\msbar{{\rm \overline{MS\kern-0.14em}\kern0.14em}}
\def\lat{{\rm lat}}
\def\glat{g_{\lat}}
\def\gbarsf{\bar{g}_{\rm SF}}

\def\alphabar{\alpha}
\def\alphasf{\alpha_{\rm SF}}
\def\alphat{\tilde{\alpha}_0}
\def\alphamsbar{\alpha_{\msbar}}

 
\def\SU{{\rm SU}(N)}
\def\SUtwo{{\rm SU}(2)}
\def\SUthree{{\rm SU}(3)}
\def\su{{\rm su}(N)}
\def\sutwo{{\rm su}(2)}
\def\suthree{{\rm su}(3)}
\def\pauli#1{\tau^{#1}}
\def\Ad{{\rm Ad}\,}

\def\Nf{N_{\rm f}}
\def\psibar{\bar{\psi}}
\def\csw{c_{\rm sw}}
\def\ct{c_{\rm t}}
\def\ctildet{\tilde{c}_{\rm t}}
\def\mc{m_{\rm c}}
\def\mr{m_{\rm R}}
\def\Zm{Z_{\rm m}}

\begin{titlepage}

\vspace*{-1.cm}
\begin{flushright} FSU-SCRI-99-71 \end{flushright}
\begin{flushright} HUB-EP-99/57 \end{flushright}
\begin{flushright} MPI-PhT/99-50\end{flushright}

\vspace{0.5cm}
\begin{center}
{\LARGE Two Loop Computation of the Schr\"odinger Functional
in Lattice QCD\\}
\vskip 1 cm
\vbox{
\centerline{
\epsfxsize=2.5 true cm
\epsfbox{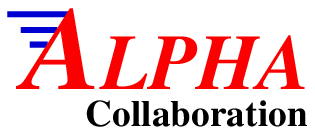}}
}
\vskip 1 cm

{Achim Bode\\
SCRI, Florida State University\\
Tallahassee, Fl 32306-4130, USA\\}
\vspace{.5cm}
{Peter Weisz\\
Max-Planck-Institut f\"ur Physik\\
F\"ohringerring 6, D-80805 M\"unchen, Germany\\}
\vspace{.5cm}
{Ulli Wolff\\
Institut f\"ur Physik, Humboldt Universit\"at\\
Invalidenstr. 110, D-10099 Berlin, Germany\\}
\end{center}
\vspace{.5cm}
\thispagestyle{empty}
\begin{abstract}\normalsize
We compute  the Schr\"odinger functional (SF) 
for the case of lattice QCD with Wilson fermions (with and
without SW improvement) at two-loop order
in lattice perturbation theory.
This allows us to extract the three-loop $\beta$-function in the SF-scheme.
These results are required to compute the running coupling,
the $\Lambda$-parameter
and quark masses by finite size techniques with negligible
{\it systematic} errors.
In addition our results enable the implementation of two-loop
O($a$) improvement in SF-simulations.

\noindent This article is based on the revised version of ref.[11].
 
\end{abstract}

\end{titlepage}

\section{Introduction}

Physical amplitudes in quantum chromodynamics (QCD)
obey renormalization group equations whose solutions
at high energy can be computed 
(in various leading logarithmic approximations) 
using renormalized perturbation theory. 
The solutions involve renormalization
group invariant mass parameters, the so-called
(scheme dependent) $\Lambda$ parameter and the 
renormalization group invariant quark masses $M$.
The determination of the relationships of these quantities 
to the low-lying hadron masses is a non-perturbative problem.  
For the quark masses there are various
non-perturbative approaches including chiral perturbation
theory, but the most systematic approach and the only
one available for computing e.g. the ratio $\Lambda/M_{\rm proton}$ 
is to employ the lattice regularization. 
There is already a vast literature on this subject 
(see e.g. the reviews \cite{rev3}-\cite{rev1}).

Among the collaborations which have as one of 
their \aend main goals the computation of these relationships
to as  high precision \aend as presently possible is the
Alpha Collaboration. The collaboration's results in the quenched
approximation have already been presented, 
most recently in \cite{CLSW}.
The extension of the project to full QCD is now under way but
due to the enormous extra cost in CPU time, 
without the sufficient compensation of computer
power it will still take some time \cite{juri} until one can match the
accuracy attained in the quenched approximation. 

The emphasis of the Alpha project is on precision and
directly associated with this, an attempt to carefully control 
systematic errors. The aim is to compute non-perturbatively defined
running couplings and quark masses 
over a wide range of energies ranging
from low energies where contact with hadronic mass parameters
is made. To make contact with perturbation theory it
is essential that the simulation reaches a range of high energy 
where the predicted behavior actually appears to set in.  
Our tactic to reach this goal  
has been explained in many publications
\cite{LNWW}, \cite{LSWW} 
and we refer the reader to these papers for further technical details. 
The key idea is to define a coupling $g(L)$ 
in a system with finite linear extension $L$ (with
specified boundary conditions). Adjusting $a/L$ and the 
bare coupling so that $g(L)$ is kept fixed,
the approach to the continuum limit of the coupling at twice
the extension $g(2L)$ depends only on $a/L$. 
As a working hypothesis this approach
is fitted by a power law (up to logarithmic factors) as expected
from the Symanzik effective action analysis \cite{rev3}.

A specific such \aend coupling $\alphasf=\gbarsf/4\pi$, measured by
the Alpha Collaboration 
is not directly experimentally accessible, but 
it can be related to phenomenological running couplings 
e.g. the $\msbar$ coupling of dimensional regularization defined
in infinite volume at high energies, using renormalized
perturbation theory:
\be
\alphamsbar(sq)=\alphasf(q)+c_1(s)\alphasf(q)^2+c_2(s)\alphasf(q)^3+\dots
\label{msbar_to_sf}
\ee
To reduce the estimated systematic
errors introduced by truncation of the perturbative series  
to the level of $\sim 1\%$ it is necessary to work out this
connection to two-loop order. For technical reasons this
is at present accomplished in two steps. The first step, the 
computation of the relationship
of $\alphamsbar$ to the bare lattice coupling $\alpha_0$, 
has been completed in refs.~\cite{LWms},\cite{CPFV} and \cite{CFPV2}.
In this paper we perform the second step,
the relationship of $\alphasf$ to $\alpha_0$ for $\Nf>0$ flavors
(the case $\Nf=0$ has already been presented in 
\cite{achimlat97}-\cite{BWWsu3YM}).

The final result between the continuum quantities 
(\ref{msbar_to_sf}) is of
course regularization independent, but the intermediate
relations in both steps 
depend on the details of the lattice action employed.
In our simulations we attempt to reduce lattice
artifacts in the continuum limit by working with the
Symanzik O$(a)$-improved fermion action of 
Sheikholeslami and Wohlert \cite{ShWo}.  
The cancellation of extra $\rmO(a)$ effects 
due to the presence of boundaries requires the
introduction of additional boundary terms in the action. 
The weights of these
terms can also be computed in perturbation theory
and the computation presented here fixes the one 
relevant for the boundary conditions under consideration
to two-loop order. 

After recalling the 
definition of $\alphasf$ in the next section
we will present some technical details, describing the 
Feynman diagrams involved in section 3.
In section 4 we discuss the 
dependence of the perturbative coefficients
on the lattice cutoff, and the related task of
determining the boundary improvement coefficients mentioned above.
In the final section 5 we compute the coefficients
in eq.(\ref{msbar_to_sf}).  
A summary of part 
of this work has already been given in \cite{achimlat99}.

\section{Basic definitions}

The Schr\"odinger functional \cite{LNWW} is the QCD partition function
\be
 {\cal Z}=e^{-\Gamma} = \int D[U]D[\psibar]D[\psi] e^{-S}\,,
\label{gammadef}
\ee
with the following particular geometry. In the spatial directions
we have a finite box of size $L$ (in all directions)
with periodic-like boundary conditions (see below). 
The extent of the time ($x_0$)
direction is also finite, $T$, and at the boundaries $x_0=0, x_0=T$,  
Dirichlet boundary conditions (specified below) are imposed.
The ratio $T/L$ belongs to the definition of the SF and in this
work we always restrict ourselves to the choice $T=L$.

The SF can be straightforwardly defined non-perturbatively by
using the lattice regularization. In eq.(\ref{gammadef})
the links interior to the box are integrated over with the
invariant SU(3) measure, and the dynamical Grassmann fields 
$\psi(x),\psibar(x)$ are those at points $x$ with $0<x_0<L$. 

The action is taken to be 
\be
S[U,\bar{\psi},\psi]=S[U]+S_{\rm W}[U,\bar{\psi},\psi]+
S_{\rm SW}[U,\bar{\psi},\psi].
\ee
The pure gauge part 
is defined by the usual sum over oriented plaquettes,
\be
S(U) = \frac1{g_0^2} \sum_p w(p) \, \tr(1-U_p).
\ee
The weight $w(p)$ is unity for all plaquettes
except those at the boundary that
contain the time-direction and one of
the frozen spatial links where we put
\be
w(p) = \ct(g_0)\,. 
\ee

$S_{\rm W}$ has the form of the standard Wilson fermion action with $r=1$,
\be
S_{\rm W}[U,\bar{\psi},\psi]=\sum_{x, \, 0<x_0<L-1}
\bar{\psi}(x)(D+M_0(x))\psi (x),
\label{swilson}
\ee
where $M_0(x)$ is the usual bare mass $m_0$, 
modified only by extra contributions at the boundaries
\be
M_0(x)=m_0+(\ctildet(g_0)-1)\Bigl[\delta_{x_0,1}+\delta_{x_0,L-1}\Bigr]\,.
\ee
The lattice Wilson Dirac operator $D$ is given by
\be
D=\frac12 \sum_{\mu}\{\gamma_{\mu}(\nabla_{\mu}^*+\nabla_{\mu})
-\nabla_{\mu}^*\nabla_{\mu}\}\,.
\ee
Hermitian $\gamma$-matrices are
used (we adopt the conventions of ref.~\cite{stefanrainer}).
$\nabla_{\mu},\nabla_{\mu}^*$ are the forward and backward lattice
covariant derivatives
\bea
\nabla_{\mu}\psi(x) &= \lambda_{\mu }U(x,\mu)\psi(x+\hat{\mu})-\psi(x),
\label{nabla1}
\\
\nabla_{\mu}^*\psi(x) &= \psi(x)-\lambda^*_{\mu }
U(x-\hat{\mu},\mu)^{\dagger}\psi(x-\hat{\mu})\,.
\label{nabla2}
\eea
Note that to be able to write the action 
(\ref{swilson}) in a more elegant form,
we have defined $\psi(x)=0=\bar{\psi} (x)$ for $x_0=0,L$.
The additional $U(1)$ field in (\ref{nabla1},\ref{nabla2}) 
\be
  \lambda_{\mu}=\cases{1      & if $\mu=0$,\cr
                      \noalign{\vskip1ex}
                      \rme^{i\theta/L} &if $\mu>0$, \cr}
\label{thetabc}
\ee
corresponds to a free phase in the spatial boundary 
conditions of the fermions. 

Finally $S_{\rm SW}$ is the Sheikholeslami-Wohlert \cite{ShWo} term
\be
S_{\rm SW}[U,\bar{\psi},\psi]
=\frac{i}{4} \csw (g_0)\sum_{x,\mu,\nu} \bar{\psi}(x)
\sigma_{\mu\nu}{\cal F}_{\mu\nu}(x)\psi(x),
\ee
where we refer to ref.~\cite{stefanrainer}
for the definition of the field ${\cal F}_{\mu\nu}$ appearing here.
The function $\csw (g_0)$ is assumed to have a
perturbative expansion of the form
\be
\csw (g_0)=\csw^{(0)}+\csw^{(1)} g_0^2 +\dots
\ee
Appropriate specification of $\csw (g_0)$
allows the elimination of all O$(a)$ effects in on shell quantities;
the values of the coefficients required for this improvement will
be denoted by $\csw^{(i)*}$.
At tree level $\csw^{(0)*}=1$ (or $=r$ with general Wilson
parameter).
The value of the 1-loop coefficient $\csw^{(1)*}$
is known (see Table \ref{pertcoeffs}) and it is 
independent of $\Nf$; we note also
that $\csw^*$ has been computed non-perturbatively for a range of bare
coupling $g_0$ in the quenched theory \cite{nonpertcswquenched1}, 
\cite{nonpertcswquenched2} and in the unquenched theory with
$\Nf=2$ \cite{nonpertcswunquenched}.

To define renormalized quark masses from the bare mass
$m_0$ for Wilson fermions requires both additive and
multiplicative renormalization:
\be
\mr=\Zm (m_0-\mc)\,.
\ee
where the critical mass $\mc$ has a perturbative expansion:
\be
\mc=\mc^{(1)}g_0^2+\mc^{(2)}g_0^4+\ldots.
\ee
The coefficients $\mc^{(i)}$ depend on $\csw$; in particular
the 1-loop coefficient $\mc^{(1)}$, which appears in our
computation, depends on 
the tree value $\csw^{(0)}$ (see Table \ref{pertcoeffs}), but is
independent of $\Nf$.

\begin{table}[ht]
\begin{center}
\begin{tabular}{|c|l|c|}\hline
\rule[-0.5ex]{0ex}{3.2ex}
 coefficient 
&value
&references\\
\hline\hline
\rule{0mm}{2.5ex}
$\mc^{(1)}|_{\csw^{(0)}=0}$&$-0.4342856(3)$&\cite{Stehr},\cite{GIM},\cite{GHS}\\
&&\\
$\mc^{(1)}|_{\csw^{(0)}=1}$&$-0.2700753495(2)$&\cite{Wohlert},\cite{LWca}\\
&&\\
\hline
\rule[-1.0ex]{0mm}{3.5ex}
$\csw^{(1)*}$&$\phantom{-}0.26590(7)$&\cite{Wohlert},\cite{LWca}\\
\hline
\end{tabular}
\caption[]{\label{pertcoeffs}
Perturbative coefficients for gauge group SU(3)}
\end{center}
\end{table}

It remains to specify the boundary conditions.
Firstly the spatial links $U(x,k), k=1,2,3$,
are frozen to fixed SU(3)-elements
for the layers at time coordinate $x_0=0,L$,
\bea
U(x,k)|_{x_0=0}&=& \exp(i a C)=
\exp\Big[\,\frac{ia}{L}\,\diag(\phi_1,\phi_2,\phi_3)\Big],
\label{surfaceB}\\
U(x,k)|_{x_0=L}&=& \exp(i a C')=
\exp\Big[\,\frac{ia}{L}\,\diag(\phi'_1,\phi'_2,\phi'_3)\Big].
\label{surfaceT}
\eea
Here $a$ is the lattice spacing, and
$\phi_i, \phi_i'$ are certain \cite{LSWW} 
dimensionless numbers that depend on
one free parameter $\eta$. 
The SF-coupling
$\gbar$ is defined from the response in 
the free energy $\Gamma$ (\ref{gammadef}),
to infinitesimal changes in the surface fields by varying $\eta$,
\be
\gbar^2 = k/\Gamma' \; ,
\ee
where $\Gamma'$ is the derivative with respect to $\eta$ at $\eta=0$,
and $k$ is a constant which is fixed by normalizing the leading term in
the perturbative expansion
\be
 \gbar^2(L) = g^2_0 + p_1(I) g^4_0 + p_2(I) g^6_0 + \ldots
\label{sf_to_bare}
\ee
where here and in the following we often denote the number of lattice
points (in a given direction) $L/a$ by $I$.

The coefficients $c_i(s)$ in the continuum 
relation (\ref{msbar_to_sf}) depend not only on the
number of quark flavors $\Nf$ but also on the particular choice
for the
boundary phases $\phi,\phi'$ occurring in (\ref{surfaceB},\ref{surfaceT})
and on the parameter $\theta$ (\ref{thetabc}).  
Both our perturbative calculation here
and simulations reported in \cite{LSWW} and more recently in
\cite{newsims,CLSW,juri} are restricted
to the choice ``A'' of \cite{LSWW} for the background field
and $\theta=\pi/5$. 

The coefficients $p_i$ appearing in
(\ref{sf_to_bare}) depend on the number of flavors,
\bea
p_1&=&p_{10} + p_{11}\Nf\,,
\label{FLAVORBREAKUP1}
\\
p_2&=&p_{20} + p_{21}\Nf + p_{22}{\Nf}^2\,. 
\label{FLAVORBREAKUP2}
\eea
The one- and two-loop coefficients $p_{10},p_{20}$ 
for the quenched case have been 
computed in refs.~\cite{LSWW},~\cite{BWWsu3YM} respectively
(and for the case of SU(2) in \cite{NW}); 
we note that in these papers
the coefficients $p_{i0}$ were denoted by $m_i$. 
The one-loop coefficient $p_{11}$ was computed 
in ref.~\cite{stefanrainer}.
Our objective here is to compute the remaining
two-loop coefficients $p_{21}, p_{22}$.

The Callan-Symanzik equation
\be
L \frac{\del}{\del L} \gbar(L) = - \beta(\gbar) = b_0 \gbar^3 
+ b_1\gbar^5
+ b_2 \gbar^7 + \ldots
\label{SFbetafunction}
\ee
(the universal 1- and 2-loop beta-function coefficients
are given in Appendix A)
requires for all values of the improvement coefficients that 
\bea
p_1 &=& 2 b_0 \ln(I) + \bar{p}_1+ \rmO(1/I) , \\
p_2-p_1^2 &=& 2 b_1 \ln(I) + \bar{p}_2 + \rmO(1/I) .
\eea
where $\bar{p}_i$ are independent of $L/a$ but dependent
on $\Nf$ and on $\csw$. The known values including our 
results\footnote{Reanalysis of the 2-loop $N_f=0$ data 
with the method of appendix D has led to errors
smaller than the very conservative estimate of \cite{BWWsu3YM}.} here 
are summarized in Table \ref{pert_coupling_coeffs}, where components
$\bar{p}_{ij}$ are defined in analogy to (\ref{FLAVORBREAKUP1}),
(\ref{FLAVORBREAKUP2}). \aend

\begin{table}[ht]
\begin{center}
\begin{tabular}{|l|l|l|c|}\hline
\rule[-1.5ex]{0mm}{3.8ex}
$(i,j)$
&$\bar{p}_{ij}|_{\csw^{(0)}=0}$
&$\bar{p}_{ij}|_{\csw^{(0)}=1}$
&references\\
\hline
\hline
$(1,0)$&$\phantom{-}0.36828215(13)
$&$\phantom{-}0.36828215(13)$&\cite{LSWW}\\
$(1,1)$&$-0.009868186(4)$&$-0.034664940(4)$&\cite{stefanrainer}\\
$(2,0)$&$\phantom{-}0.048091(2)
$&$\phantom{-}0.048091(2)$&\cite{BWWsu3YM}\\
$(2,1)$&$-0.00349(3)$&$\phantom{-}0.00978(2)-0.054641(1)\csw^{(1)}$&\\
$(2,2)$&$\phantom{-}0.000211(3)$&$\phantom{-}0.000209(1)$&\\
\hline
\end{tabular}
\caption[]{\label{pert_coupling_coeffs}
SF coupling coefficients for gauge group SU(3), background field A
and $\theta=\pi/5$}
\end{center}
\end{table}

The boundary weight $\ct$ 
is assumed to have a perturbative expansion of the form
\be
\ct(g_0) = 1 + \ct^{(1)}g_0^2 + \ct^{(2)}g_0^4 + \ldots
\ee
and similarly for $\ctildet$.
The freedom of adjusting the boundary weights $\ct$ and $\ctildet$ 
to specific functions $\ct^*$ and $\ctildet^*$ is
required for improvement of $\rmO(a)$ lattice artifacts
that are otherwise introduced by the surfaces. 
The one-loop coefficient $\ctildet^{(1)*}$ required for improvement 
of SF correlation functions has been computed in ref.~\cite{LWca}
and is independent of $\Nf$.
With this knowledge, the perturbative coefficients 
of $\ct^*$ can be computed to 2-loops 
from evaluating the $\rmO(a)$ lattice artifacts of
the SF coupling and demanding deviations from leading
continuum behavior to be of $\rmO(a^2)$,  
\bea
\Bigl[ p_1 \Bigr]_{c=c^*} &=& 2 b_0 \ln(I) + \bar{p}_1+ \rmO(1/I^2)\,,
\\
\Bigl[ p_2 - p_1^2 \Bigr]_{c=c^*}
&=& 2 b_1 \ln(I) + \bar{p}_2 + \rmO(1/I^2)\,.
\eea
The coefficients
$\ct^{(i)*}$ do depend \aend on the number of fermions,
\bea
  \ct^{(1)*}&=& \ct^{(1,0)*}+\ct^{(1,1)*}\Nf\,, \\
  \ct^{(2)*}&=& \ct^{(2,0)*}+\ct^{(2,1)*}\Nf+\ct^{(2,2)*}{\Nf}^2\,. 
\eea
The values\footnote{See previous footnote.} 
of these coefficients are given in Table \ref{pert_bdy_coeffs}.

\begin{table}[ht]
\begin{center}
\begin{tabular}{|c|l|c|}\hline
\rule[-0.8ex]{0mm}{3.2ex}
 coefficient 
& $\phantom{-}$ value
&references\\\hline
\rule[-1.2ex]{0mm}{3.8ex}
$\ctildet^{(1)*}$&$-0.01795(2)$&\cite{LWca},\cite{pwss}\\
\hline
\rule[-1.0ex]{0mm}{3.8ex}
$\ct^{(1,0)*}$&$-0.08900(5)$&\cite{LNWW}\\
\rule[-1.0ex]{0mm}{3.8ex}
$\ct^{(1,1)*}$&$\phantom{-}0.0191410(1)$&\cite{stefanrainer}\\
\rule[-1.0ex]{0mm}{3.8ex}
$\ct^{(2,0)*}$&$-0.0294(3)$&\cite{BWWsu3YM}\\
\rule[-1.0ex]{0mm}{3.8ex}
$\ct^{(2,1)*}$&$\phantom{-}0.002(1)$&\\
\rule[-1.0ex]{0mm}{3.8ex}
$\ct^{(2,2)*}$&$\phantom{-}0.0000(1)$&\\
\hline
\end{tabular}
\caption[]{\label{pert_bdy_coeffs}
SF boundary $\rmO(a)$ improvement coefficients for gauge group SU(3)}
\end{center}
\end{table}

\section{Perturbation expansion to two loops}

The perturbative expansion in the bare coupling amounts to an
expansion of the gauge field around the induced background field.
This has been described in detail in previous publications
~\cite{LNWW},~\cite{NW},~\cite{BWWsu3YM}.
In particular we use the gauge fixing procedure as in ref.~\cite{LNWW}.
In the presence of the abelian background field, both the
gluon and quark propagators cannot be computed analytically.
The numerical computation of the gluon propagator is described
in Section 2 of ref.~\cite{BWWsu3YM}, 
and here in Appendix B we give some technical details 
on how we numerically computed the fermion propagator ${\cal S}$.

An enumeration of the Feynman diagrams 
contributing to the Schr\"{o}dinger functional 
to two-loop order
reveals that the coefficients $p_{ij}$ depend on the
improvement coefficients and critical mass in the following way:
\bea
  p_{10} &=& p_1^a + \ct^{(1,0)} p_1^b\,, \\
  p_{11} &=& p_1^c + \ct^{(1,1)} p_1^b\,, \\
  p_{20} - p_{10}^2 &=& p_2^a + \ct^{(1,0)} p_2^b
  + \left[\ct^{(1,0)}\right]^2 p_2^c + \ct^{(2,0)} p_2^d\,,\\
  p_{21} - 2p_{10}p_{11}
  &=& p_2^e + \ct^{(1,1)} p_2^b
  + 2\ct^{(1,0)}\ct^{(1,1)} p_2^c + \ct^{(2,1)} p_2^d \nonumber\\
  && + \ct^{(1,0)}p_2^f+ \ctildet^{(1)}p_2^g+\csw^{(1)}p_2^h
  +\mc^{(1)}p_2^i\,,\\
  p_{22} - p_{11}^2 &=& p_2^j
  + \left[\ct^{(1,1)}\right]^2 p_2^c + \ct^{(2,2)} p_2^d
  +\ct^{(1,1)}p_2^f\,.
  \label{impabcd}
\eea

All the coefficients coming from diagrams not involving the quarks
have been computed in \cite{BWWsu3YM}. Here we recall that only
two of the coefficients can be given in closed form:
\bea
p_1^b&=& p_2^d = -\frac{2}{I}\,,
\\
p_2^c&=& \frac{2}{I} -\frac{4}{I^2}+ \rmO(1/I^5)\,,
\eea
whereas the others require careful numerical evaluation.
The contributions $p_1^a$, $p_2^a$ and $p_2^b$ are tabulated
in Table 1 of ref.~\cite{BWWsu3YM}.

As for the diagrams involving the quarks, the one-loop coefficient
\be
p_1^c=-{1\over k}\Gamma_1^{c\prime}\,,
\ee
with
\be        
\Gamma_1^{c\prime}=-{\rm tr} \,\,{\cal S} 
{\partial {\cal S}^{-1}\over\partial\eta}|_{\eta=0}\,, 
\ee 
has been computed in  ref.\cite{stefanrainer}.

It remains to compute the
coefficients $p_2^e,p_2^f,p_2^g,p_2^h,p_2^i,p_2^j$.
The contributions $p_2^g,p_2^h,p_2^i$ are very simple because
they are obtained by differentiating the one-loop
contribution $p_1^c$ with respect to the bare parameters e.g.
\be 
\Gamma_2^{h\prime}={\rm tr} \Bigl[
\,\,{\cal S}{\partial {\cal S}^{-1}\over\partial\eta}
{\cal S}{\partial {\cal S}^{-1}\over\partial\csw(0)}
-{\cal S}{\partial^2 {\cal S}^{-1}\over\partial\eta\partial\csw(0)}
\Bigr]|_{\eta=0}\,,
\ee
and similarly for $p_2^g,p_2^i$. The contributions are diagrammatically
represented in Fig. 1a where the filled circle indicates the insertion
obtained by differentiation with respect to one of the bare parameters
(the differentiation with respect to $\eta$ is not indicated). 

The contribution $p_2^f$ arises from the correlation of a boundary term
in the pure gluon part of the action with the quark-antiquark 1-gluon
vertex given in diagram Fig. 1b.

\hspace{1cm}
\begin{picture}(170,40)(0,17)
\SetOffset(30,0)
\ArrowArc(40,20)(18,270,630)
\Vertex (40, 2){3}
\end{picture} 

\vspace{-1.4cm}
\hspace{7cm}
\begin{picture}(170,40)(0,17)
\ArrowArc(68,20)(18,180,540)
\Gluon  (10,20)(50,20){4}{4}
\GCirc(10,20){2}{0.7}
\Vertex (50,20){2}
\end{picture}

\vspace{1cm}
\hspace{2.5cm} Fig. 1a \hspace{4.0cm} Fig. 1b
\vspace{0.5cm}

The remaining part $p_2^e$ of $p_{21}$ comes from the four diagrams
in Fig. 2. The fermion big-mac diagram Fig. 2c is 
the technically most difficult diagram; 
the rest of the diagrams are simpler
because they are essentially products of 1-loop diagrams.   

\begin{picture}(170,40)(0,17)
\SetOffset(30,0)
\ArrowArc(85,20)(18,180,540)
\GlueArc (15,20)(15,0,360){3}{9}
\Gluon  (33,20)(67,20){4}{4}
\Vertex (33,20){2}
\Vertex (67,20){2}
\end{picture} 

\vspace{-1.4cm}
\hspace{6cm}   
\begin{picture}(170,40)(0,17)
\SetOffset(30,0)
\DashArrowArc(15,20)(18,  0,360){4}
\ArrowArc(85,20)(18,180,540)
\Gluon  (33,20)(67,20){4}{4}
\Vertex (33,20){2}
\Vertex (67,20){2}
\end{picture} 

\vspace{1cm}
\hspace{2.0cm} Fig. 2a \hspace{4.5cm} Fig. 2b
\vspace{1cm}

\begin{picture}(170,40)(0,17)
\SetOffset(50,0)
\ArrowArc(30,20)(20,  0,180)
\ArrowArc(30,20)(20,180,360)
\Vertex (10,20){2}
\Vertex (50,20){2}
\Gluon  (10,20)(50,20){4}{4}
\end{picture} 

\vspace{-1.4cm}
\hspace{6cm}   
\begin{picture}(170,40)(0,17)
\SetOffset(33,0)
\ArrowArc(30,20)(18,  0,360)
\GlueArc     (63,20)(15,180,540){3}{9}
\Vertex      (48,20){2}
\end{picture} 

\vspace{1cm}
\hspace{2.0cm} Fig. 2c \hspace{4.5cm} Fig. 2d
\vspace{0.5cm}

Finally there is only one as yet unmentioned contribution 
$p_2^j$ to $p_{22}$, which is basically the product of two
fermion tadpoles depicted in Fig. 3.

\hspace{3cm}
\begin{picture}(170,40)(0,17)
\SetOffset(30,0)
\ArrowArc(15,20)(18,  0,360)
\ArrowArc(85,20)(18,180,540)
\Gluon  (33,20)(67,20){4}{4}
\Vertex (33,20){2}
\Vertex (67,20){2}
\end{picture}
\vspace{1cm}

\hspace{5.0cm} Fig. 3

\vspace{1cm}

Despite optimization of the code in various respects by e.g.
making use of all symmetry properties, the CPU-time for the computation 
which is dominated by \aend the big-mac diagram in Fig. 2c,
is rather costly growing rapidly as $(L/a)^5$. All diagrams were
computed using double precision arithmetic to at least $L/a=32$, 
for two cases $\csw^{(0)}=1$ and $\csw^{(0)}=0$. 
The numerical results for $p_2^e,p_2^f,p_2^g,p_2^h,p_2^i,p_2^j$ 
for $\csw^{(0)}=1$, and for $p_2^e,p_2^i,p_2^j$
for $\csw^{(0)}=0$,
are given in Appendix C in Tables \ref{ptable1},\ref{ptable2}
and \ref{ptable3} respectively. 

We have various stringent tests (as in \cite{NW}) to check our
numbers;
firstly the full code using sums in position space was compared
on the smallest lattices
with a slow but independent code (many of whose subroutines had been
well tested in previous computations) making sums in momentum space. 
Secondly gauge parameter independence of the $p_2^x$ was checked.
Thirdly it was checked that the full code 
reproduced numbers for partial sums satisfying all analytically known
symmetries, before these were actually used to reduce the number of terms
in order to speed up the program execution time.
Finally the fact that the $I$-dependence for 
$p_{21},p_{22}$ (depending on the value of $\csw^{(0)}$
as described in the next section) was as expected from general
considerations, gave a further consistency check.

\section{Analysis of the $a/L$ dependence}

The generically expected behavior for 
2-loop lattice Feynman diagrams suggests an asymptotic expansion
for the two-loop coefficients $p_2^x$ of the form (recall $I=L/a$)
\be
p_2^x = \sum_{n=n^x}^{\infty} [r_n^x+s_n^x \ln(I) 
+ t_n^x\ln^2(I)]I^{-n}.
\ee
The coefficients of these expansions are extracted 
from the series of finite lattices (see Appendix C). The method 
we used here to extract these numbers together with an 
estimate of their systematic errors is described in
some detail in Appendix D.

\subsection{The case $\csw^{(0)}=1$}

\subsubsection{The continuum behavior of $p_{22}$}

Let us start with the simpler case of $p_{22}$ where the non-trivial
contributions come from only $p_2^f$ and $p_2^j$.
For $p_2^f$ we have $n^f=1$ and $t_n^f=0$ for all $n$.
Our result is:
\bea
s_1^f&=&-0.03375(4)\,,
\\
r_1^f&=&-0.0734(4)\,.
\eea
The contribution $p_2^j$ on the other hand has a finite non-trivial
continuum limit $n^j=0,\,\,s_0^j=0$. As for the coefficients 
$t_n^j$ these are generally non-zero, since the diagram is a
product of one-loop diagrams, but $t_0^j=t_1^j=0$.    
For the leading coefficients we find
\bea
r_0^j&=&0.000210(3)\,,
\\
s_1^j&=&0.0003(14)\,,
\\
r_1^j&=&0.003(9)\,.
\eea
The extracted value of $s_1^j$ above is unfortunately not very precise
but consistent with the value $0.000646$ obtained from the equation
\be
s_1^j+\ct^{(1,1)*}s_1^f=0
\label{impcond1}
\ee
that we expect from $\rmO(a)$ improvement.

If we now analyze the series for $p_2^j+\ct^{(1,1)*}p_2^f$ \aend
with the known value for $\ct^{(1,1)*}$ with the assumption that the
equation (\ref{impcond1}) is fulfilled i.e.
\be
p_2^j+\ct^{(1,1)*}p_2^f=r_0^{fj}+{r_1^{fj}\over I}
+{r_2^{fj}+s_2^{fj}\ln(I)+t_2^{fj}\ln^2(I)\over I^2}+\dots
\ee
we get
\bea
r_0^{fj}&=&\phantom{-}0.000209(1)\,,
\\ 
r_1^{fj}&=&-0.0008(2)\,.
\eea
Thus we obtain a refined estimate of $\bar{p}_{22}=r_0^{fj}$ 
recorded in Table \ref{pert_coupling_coeffs}. 
Finally we estimate the 
improvement coefficient $\ct^{(2,2)*}$ through 
\be
\ct^{(2,2)*}=\frac12 r_1^{fj}+\Bigl(\ct^{(1,1)*}\Bigr)^2=0.0000(1)\,.
\ee

\subsubsection{The continuum behavior of $p_{21}$} 

First of all, the expansion of $p_2^g$ has the same structure
as that of $p_2^f$ i.e. it has a trivial continuum limit $n^g=1$
and also the coefficients $t_n^g=0$ for all $n$. The analysis yields
\bea
s_1^{g}&=&0.0338(3)\,,
\\
r_1^{g}&=&0.115(2)\,.
\eea
$p_2^h$ on the other hand has a non-trivial finite continuum
limit $n^h=0,\,s_0^h=0$ and $t_n^h=0$ for all $n$. Our results
for the leading terms are
\bea
r_0^{h}&=&-0.054641(1)\,,
\\
s_1^{h}&=&-0.0252(2)\,,
\\
r_1^{h}&=&\phantom{-}0.032(1)\,.
\eea
The mass insertion term $p_2^i$ has a linear divergence,
$n^i=-1$ but $s_{-1}^i=s_0^i=0$ and  $t_n^i=0$ for all $n$.
For the leading terms we find
\bea
r_{-1}^{i}&=&0.009568(1)\,,
\\
r_0^{i}&=&0.01198(11)\,.
\eea
Finally the contribution $p_2^e$ has an expansion of the form
\be
p_1^e=r_{-1}^eI+r_0^e+s_0^e\ln(I)+
{r_1^e+s_1^e\ln(I)\over I}+\dots
\ee
$t_1^e$ is expected to be zero because of tree level improvement
and the data is consistent with this expectation. 
For this coefficient
our analysis gives \aend
\bea
r_{-1}^e &=& \phantom{-}0.002586(2)\,, \\
s_0^e    &=&           -0.0012(3)\,, \\
r_0^e    &=& \phantom{-}0.0141(16)\,.
\eea
On the other hand we know that the linear divergences must cancel
and hence 
\be
r_{-1}^e=-\mc^{(1)}r_{-1}^i\,,      
\label{lindiv}
\ee
should hold. Indeed from (\ref{lindiv}) we find the value $0.0025841(3)$
which serves as an additional consistency check. 
In the next step we thus \aend
assume the cancellation of linear divergences and analyze
\be
p_2^e+\mc^{(1)}p_2^i=r_0^{ei}+s_0^{ei}\ln(I)
+{r_1^{ei}+s_1^{ei}\ln(I)+t_1^{ei}\ln^2(I)\over I}+\dots
\ee
From this series we could get the still somewhat rough estimates
of the continuum behavior
\bea
s_0^{ei}&=&-0.00104(12)\,,
\label{b11est}
\\
r_0^{ei}&=&\phantom{-}0.0099(7)\,.
\eea
The Callan-Symanzik equation
(\ref{SFbetafunction}) (which incidentally requires 
$t_0=0$) requires the logarithmic divergence
$s_0^{ei}$ to coincide with $2b_{11}=-0.0010159$, and our estimate
(\ref{b11est}) agrees with this within errors. We are thus justified 
to assume this to actually be the case and continue to analyze
\be 
p_2^e+\mc^{(1)}p_2^i-2b_{11}\ln(I)=r_0^{ei}
+{r_1^{ei}+s_1^{ei}\ln(I)+t_1^{ei}\ln^2(I)\over I}+\dots
\ee
to obtain an improved estimate for the coefficient  
\be
r_0^{ei}=0.00978(2)\,.              
\ee
The value for the coefficient $\bar{p}_{21}$ in Table
\ref{pert_coupling_coeffs} comes from
\be
\bar{p}_{21}=r_0^{ei}+\csw^{(1)*}r_0^h\,.
\ee
We also obtain (unfortunately rather poor) estimates for the 
$\rmO(a)$ coefficients
\bea
s_1^{ei}&=&-0.0016(17)\,,
\\
r_1^{ei}&=&-0.003(7)\,.
\eea
In fact improvement requires 
\be
-s_1^{ei}=\ct^{(1,1)*}s_1^b+\ct^{(1,0)*}s_1^f+\ctildet^{(1)*}s_1^g
+\csw^{(1)*}s_1^h\,.
\ee 
Using the previously obtained \cite{phd} values 
$r_1^b=0.16831(84)$ and $s_1^b=0.27848(40)$ gives
$s_1^{ei}=-0.00103(7)$ which is consistent with but much more accurate
than our estimate above.

Finally the most accurate value we could obtain for improvement
coefficient $\ct^{(2,1)*}$ was extracted by forming the combination
\be
p_2^e+\mc^{(1)}p_2^i-2b_{11}\ln(I)+
\ct^{(1,1)*}p_2^b+\ct^{(1,0)*}p_2^f+\ctildet^{(1)*}p_2^g
+\csw^{(1)*}p_2^h\,,
\ee
and analyzing the series with the assumption it has the form
\be
r_0^x+{r_1^x\over I}+{r_2^x+s_2^x\ln(I)+t_2^x\ln^2(I)\over I^2}+\dots
\ee
thereby obtaining the estimate
\be
r_1^x=0.011(2)\,,
\ee
and finally 
\be
\ct^{(2,1)*}=2\ct^{(1,0)*}\ct^{(1,1)*}+\frac12 r_1^x=0.002(1)\,.
\ee

\subsection{The case $\csw^{(0)}=0$}

The analysis of the data for the case $\csw^{(0)}=0$
is completely analogous to that for $\csw^{(0)}=1$ above,
except for the fact that we can simply ignore also all other
improvement coefficients. 
Thus e.g. for $p_{22}$ only $p_2^j$ has
to be analyzed, and with the same ansatz as before we find
\be
r_0^j=0.000211(3)\,.
\ee

For $p_{21}$ we only take into account $p_2^e$ and $p_2^i$.
Again we find that the linear divergences cancel (with the
appropriate value of $\mc^{(1)}$) and the
coefficient of $\ln(I)$ reproduces $2b_{11}$ within errors.
Finally we fit 
\be
p_2^e+\mc^{(1)}p_2^i-2b_{11}\ln(I)=r_0^x
+{r_1^x+s_1^x\ln(I)+t_1^x\ln^2(I)\over I}+\dots ,
\ee 
and the ensuing estimate of $r_0^x$ gives us the
value of $\bar{p}_{21}$ quoted in Table \ref{pert_coupling_coeffs} 
for this case.

\section{Applications}

\subsection{The relation of $\alphamsbar$ to $\alphasf$}

To obtain the coefficients $c_i(s)$ in 
eq.(\ref{msbar_to_sf}) from our computation above,
we have to use the known results for the coefficients
appearing in the relation between $\alphamsbar$ and
$\alpha_0$:
\be
\alphamsbar(s/a)=\alpha_0+d_1(s)\alpha_0^2+d_2(s)\alpha_0^3+\dots
\ee
with
\bea
d_1(s)&=&\sum_{r=0}^1 {\Nf}^r\Bigl\{ -8\pi b_{0r}\ln(s)+d_{1r}\Bigr\}\,, 
\label{d1}\\
d_2(s)&=&d_1(s)^2+\sum_{r=0}^1
{\Nf}^r \Bigl\{-32\pi^2 b_{1r}\ln(s)+d_{2r}\Bigr\}\,.
\label{d2}
\eea
The coefficients $d_{ir}$  (and the $\beta-$function
coefficients) are given in Appendix A.   

The coefficients $c_i(s)$ in eq.(\ref{msbar_to_sf}) 
are now obtained through
\bea
c_1(s)&=&-8\pi b_0\ln(s)+
\sum_{r=0}^1 {\Nf}^r\Bigl\{d_{1r}-4\pi \bar{p}_{1r}\Bigr\}\,,
\\
c_2(s)&=&c_1(s)^2-32\pi^2 b_1\ln(s)+\sum_{r=0}^2
{\Nf}^r\Bigl\{d_{2r}-16\pi^2\bar{p}_{2r}\Bigr\}\,. 
\label{c2} 
\eea
Now it is clear that since eq.(\ref{msbar_to_sf}) is a relation 
between continuum quantities
all coefficients $c_r$ must be independent of the lattice 
bare parameters,
which serves as a further consistency check on the computations.
This has already been observed at the 1-loop level
i.e. from Table \ref{pert_coupling_coeffs}
and eqns.(\ref{d11},\ref{K10}) one sees that $d_{11}-4\pi\bar{p}_{11}$
is independent of $\csw^{(0)}$. At the two loop level,
since $d_{22}=0$ (there is no term 
$\propto {\Nf}^2$ in (\ref{d2})),
we should find that $\bar{p}_{22}$ is independent of $\csw$.
Indeed our numerical results in Table \ref{pert_coupling_coeffs} are
consistent with this expectation. 
Finally $d_{21}-16\pi^2\bar{p}_{21}$
should be independent of $\csw$; unfortunately this cannot be
checked at present because the computation of the $d_{21}$ 
for $\csw\ne0$ is not yet complete \cite{HP}. 
However demanding the equality would require
\be
d_{21}|_{N=3,\,\csw^{(0)}=1}=1.685(9)-8.6286(2)\csw^{(1)}\,,
\ee
which will serve as a good consistency check 
for the (general SU($N$)) computation above in progress \cite{HP}.
In fact the coefficient of $\csw^{(1)}$ above, $16\pi^2 r_0^h$,
is already independently checked because $r_0^h$ should be 
related to the coefficient $K_1$ in (\ref{K10}) through  
\be
r_0^h={1\over4\pi} \left. {\rmd K_1(x)\over \rmd x}\right|_{x=1}\,,
\ee 
and this is fulfilled numerically to good precision.

Putting all our numerical results together we find for $N=3$
\bea
c_1(s)&=&-8\pi b_0\ln(s)+1.255621(2)+0.0398629(2)\Nf\,,
\\
c_2(s)&=&c_1(s)^2-32\pi^2b_1\ln(s)
\nonumber\\
&& +1.197(10)+0.140(6)\Nf-0.0330(2){\Nf}^2\,.
\eea 
This two-loop connection between the two different
couplings determines the difference
between the non-universal three-loop coefficients of their 
respective $\beta$-functions:
\be
b_2=b_2^{\msbar}+{b_1c_1(1)\over 4\pi}-
{b_0\Bigl[ c_2(1)-c_1(1)^2\Bigr]\over 16\pi^2}\,.
\ee
Since the three-loop $\beta$-function in the $\overline{\rm MS}$-scheme 
is known we can obtain the SF 3-loop beta function coefficient for the
background field A (and $\theta=\pi/5$) e.g. :
\bea
b_2|_{\Nf=0}&=&0.482(7)\times (4\pi)^{-3}\,,
\\
b_2|_{\Nf=2}&=&0.064(10)\times (4\pi)^{-3}\,.
\eea

The perturbative coefficients
will find their application when the SF-coupling has been
measured over a wide range of energies.
Once the SF-coupling $\gbar$ is known for small box size, 
it is to be converted
to the $\overline{\rm MS}$-coupling at high energy. One could 
conventionally choose the mass of the neutral 
weak boson $M_Z$ as a scale here.
Another procedure --- attractive for asymptotically free theories ---
is to extract the $\Lambda$-parameter which is simply related to the behavior
at asymptotically large energy.
It is a renormalization group invariant given by
\be
\Lambda_{\rm SF} = L^{-1} (b_0 \gbar^2)^{-b_1/(2b_0^2)}
\rme^{-1/(2b_0\gbar^2)}
\exp\left\{ -\int_0^{\gbar} \rmd g \left[ \frac1{\beta(g)}
+\frac1{b_0 g^3} -\frac{b_1}{b_0^2 g}
\right]\right\}.
\label{LambdaSF}
\ee
The conversion to $\Lambda_{\msbar}$ then amounts to an 
additional known factor:
\bea
\Lambda_{\msbar}&=&\Lambda_{\rm SF}
\exp \left({c_1(1)\over 8\pi b_0} \right)
\\
&=& 2.382035(3) \Lambda_{\rm SF}\,,\,\,\,{\rm for}\,\,\Nf=2\,.
\eea
If we insert a small $\gbar(L)$ belonging to 
a very small $L$ (in physical units) \aend into formula (\ref{LambdaSF}),
then the exponentiated integral is close to unity. 
Knowledge of the three loop term will give some estimate of the 
systematic error which will definitely be less than the
statistical error in the near future.

\subsection{The step scaling function}
A central quantity in the ALPHA collaboration's approach is the step
scaling function which generalizes the $\beta$-function to finite
rescalings,
\begin{eqnarray}
        \sigma(s,u)&=&\gbar^2(sL)|_{\gbar^2(L)=u}\,.
\end{eqnarray}
We remind the reader that we use a mass independent renormalization scheme
and set the quark mass to zero. \aend
On the lattice, $\sigma$ emerges as the continuum limit of a finite lattice
spacing approximant $\Sigma$,
\begin{eqnarray}
\sigma(s,u)&=&\lim_{a\rightarrow0}\Sigma(s,u,a/L).
\end{eqnarray}
Due to the absence of chiral symmetry before the continuum limit is taken
one has to precisely specify a zero mass condition with the cutoff
in place. In the numerical simulations a certain unambiguous 
definition based on the \aend
PCAC relation is adopted. This leads to $I=L/a$ dependent expansion
coefficients $m_c^{({i})}(I)$, which only as $I\to\infty$ go over to the
convention independent values $m_c^{({0})}=0$ and $m_c^{({1})}$ 
in Table~\ref{pertcoeffs}.
Only the latter are presently known to us. For the extrapolations
in the previous section (including O($a$) improvement) this is of no concern.
The following perturbative estimation of the full 
(all orders in $a$) lattice artifacts
would however be more realistic with finite $I$ mass expansion coefficients
and is hence only given to get a first idea here. An improved version will
be published elsewhere.

All perturbative information about the convergence speed 
of $\Sigma$ for $s=2$ is
conveniently summarized in coefficients $\delta_{nj}(a/L)$ defined by
\begin{eqnarray}
\delta(u,a/L)&=&
\frac{\Sigma(2,u,a/L)-\sigma(2,u)}{\sigma(2,u)}\\
&=&\sum_{n=1} u^n \sum_{j=0}^{n}
       N_{\rm f}^j\delta_{nj}(a/L) .
\end{eqnarray}
Setting $\triangle p_{ij}(I) = p_{ij}(2I)-p_{ij}(I)$ we have
\begin{eqnarray}
\delta_{10}&=&\triangle p_{10}-2b_{00} \ln(2) ,\\
\delta_{11}&=&\triangle p_{11}-2b_{01} \ln(2) ,\\
\delta_{20}&=&\triangle p_{20}-2b_{10} \ln(2) 
     -2\triangle p_{10}(p_{10}+b_{00} \ln(2))
            ,\\
\delta_{21}&=&\triangle p_{21}-2b_{11} \ln(2) 
     -2\triangle p_{11}(p_{10}+b_{00} \ln(2)) \nonumber\\
&\phantom{=}&
     -2\triangle p_{10}(p_{11}+b_{01} \ln(2))
            ,\\
\delta_{22}&=&\triangle p_{22} 
     -2\triangle p_{11}(p_{11}+b_{01} \ln(2))
            .
\end{eqnarray}

All the $\delta_{nj}$ decay with an asymptotic
rate proportional to $(a/L)^2$ in
the $O(a)$ improved theory.
We are free to use $\delta$ to cancel perturbative finite
$a$ effects from Monte Carlo data,
as pointed out in \cite{DFGLPSWW}. 
For this purpose one uses the same
improvement coefficients as in the simulation and the expansion
of $m_c$ for finite lattice spacing.  
Only for cases where 
nonperturbative improvement coefficients are used in the simulation, like
$\csw$, one replaces these by the leading perturbative expressions. 
\begin{figure}[htb]
\hspace*{0.8cm}
\epsfig{file=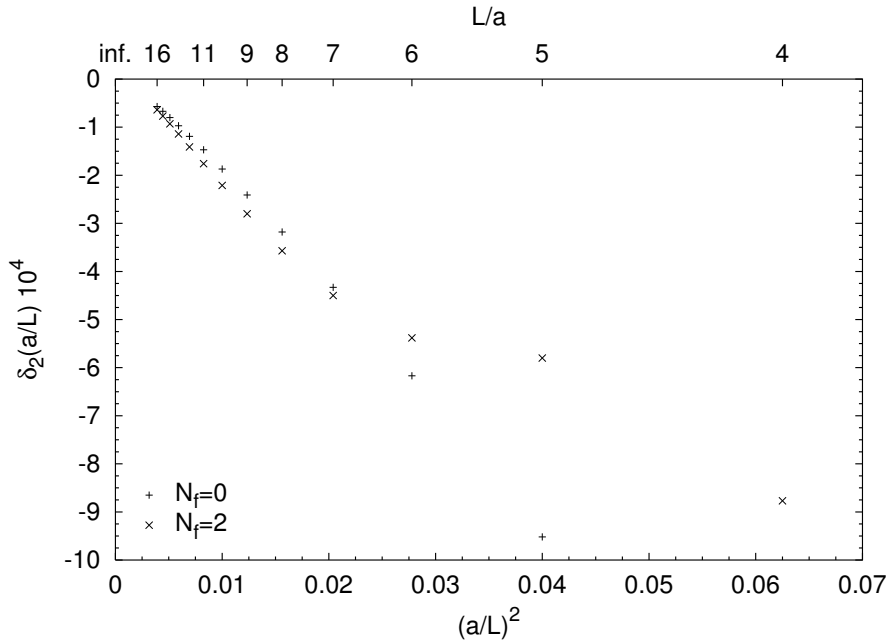,width=12cm}
\caption{\label{fig1} Two loop contribution to lattice artifacts.}
\end{figure}
Fig.\ref{fig1} gives an impression of the two loop artifacts
for zero and two flavours. They are
reasonably small and show the expected decay pattern.

{\bf Acknowledgements}

We would like to thank Juri Rolf, Stefan Sint and Rainer Sommer
for useful discussions and correspondence.


\appendix
\renewcommand{\thesection}{A. The 2-loop relation 
between $\alphamsbar$ and $\alpha_0$}
\section{}
\renewcommand{\thesection}{A}

The coefficients $d_{ir}$ appearing in 
the the relation between $\alphamsbar$ and
$\alpha_0$ Eqs.(\ref{d1}), (\ref{d2})  
have the form for general gauge group SU($N$)
\bea
d_{10}&=&-{\pi\over 2N} +k_1N\,,
\\
d_{11}&=&K_1\,,
\label{d11}
\\
d_{20}&=&{3\pi^2\over 8N^2} +k_2+k_3N^2\,,
\\
d_{21}&=&{K_2\over N}+K_3N\,.
\label{d21}
\eea
The coefficients $k_i$ which have been computed in \cite{LWms}
and checked by the authors of ref.~\cite{CPFV} are given by
\bea
k_1&=&\phantom{-}2.135730074078457(2)\,,
\label{k1}\\
k_2&=&-2.8626215972(6)\,,
\label{k2}\\
k_3&=&\phantom{-}1.24911585(3)\,.
\label{k3}
\eea

The coefficient $K_1$ is a function of $\csw^{(0)}$ 
and Wilson's parameter $r$. In our computations we
always set $r=1$ and hence only indicate the dependence on 
$\csw^{(0)}$ i.e. $K_1=K_1(\csw^{(0)})$\,.
We have 
\footnote{To give a standard reference to the computation 
of $K_1$ is difficult. The precise result in (\ref{K10})
comes from Stefan Sint \cite{sint}. The result 
in eq.(29) of the recent paper \cite{MaSa} agrees with this .
The computation of $K_1(0)$ was first presented in refs.
~\cite{KNS}, \cite{peter} and a more precise value is given 
in \cite{CFPV2}.
The extension of the calculation to $\csw\ne0$ was first
presented in \cite{CaRo}, however there are some errors in the
result quoted there (of which the authors were conscious 
but unfortunately they did not publish an erratum).} 
\be 
K_1(x)=-0.08414443(8)+0.063419(1)x-0.375024(1)x^2\,.
\label{K10}
\ee
Similarly we have $K_r=K_r(\csw^{(0)},\csw^{(1)})$ for $r=2,3$.
So far only the pure Wilson case has been computed in ref.~\cite{CFPV2} 
\bea 
K_2(0,0)&=&\phantom{-}0.1890(2)\,,
\label{K2}\\
K_3(0,0)&=&-0.1579(3)\,.
\label{K3}
\eea
The extension of the computation to the case $\csw\ne0$ is 
now under way \cite{HP}.

The universal coefficients of the beta-function are given by
\be
b_i=b_{i0}+b_{i1}\Nf,\,\,\,\,i=0,1,
\ee
with
\bea
b_{00}&=&\phantom{-}{1\over (4\pi)^2}{11 N\over3}\,,  
\label{b00}\\
b_{01}&=&-{1\over (4\pi)^2}{2\over3}\,,
\label{b01}\\
b_{10}&=&\phantom{-}{1\over (4\pi)^4}{34 N^2\over3}\,,
\label{b10}\\
b_{11}&=&-{1\over (4\pi)^4}\Bigl({13 N\over3}-{1\over N}\Bigr)\,,
\label{b11}
\eea
and the 3-loop beta function coefficient in the MS-scheme \cite{3loopMS}
is given by
\be 
b_2^{\msbar}=b_{20}^{\msbar}+b_{21}^{\msbar}\Nf+b_{22}^{\msbar}{\Nf}^2\,,
\ee
with
\bea
b_{20}^{\msbar}&=&\phantom{-}{1\over (4\pi)^6}{2857 N^3\over54}\,,
\label{b20}\\
b_{21}^{\msbar}&=&-{1\over (4\pi)^6}
\Bigl({1709N^2\over54}-{187\over36}-{1\over4N^2}\Bigr)\,,
\label{b21}\\
b_{22}^{\msbar}&=&\phantom{-}{1\over (4\pi)^6}
\Bigl({56N\over27}-{11\over18N}\Bigr)\,.
\label{b22}
\eea


\appendix
\renewcommand{\thesection}{B.}
\section{The free fermion propagator
in the abelian background field}
\renewcommand{\thesection}{B}

The Wilson-Dirac operator $D$,
which is now taken to include the Sheikoleslami-Wohlert term,
is diagonal in momentum and color space for the background field under
consideration. We hence assume these fixed 
and discuss the inversion in time and Dirac-spin space.
%
The calculational technique is adapted 
from the calculation the background field gluon propagator
in \cite{NW,BWWsu3YM}.

The propagator will be entirely
constructed from solutions to the homogeneous
equations
\begin{eqnarray}
(D\psi^f)(t)=0 && \mbox{ with } P_+ \psi^f(0) = \psi_+^f(0) =0,\\
(D\psi^b)(t)=0 && \mbox{ with } P_- \psi^b(T) = \psi_-^b(T) =0
\end{eqnarray}
with projectors $P_{\pm} = \frac12 (1 \pm \gamma_0)$.
In each case there are two independent solutions, which are
constructed by forward (backward) recurrence in $t$ for $\psi_f$ ($\psi_b$)
starting from the $t=0$ ($t=T$)
boundary. We assemble the two solutions into two columns in $\psi_f$ , $\psi_b$.
Using pseudo hermiticity, $D^{\dagger} =\gamma_5 D \gamma_5$,
the defining equation
\begin{equation}
D {\cal S}(t,t') = \delta_{t,t'}, \; 0<t,t'<T \, ,
\label{propeq}
\end{equation}
and the fact that in a matrix sense $S$ must also be the left inverse to $D$
we derive
\begin{eqnarray}
{\cal S}(t,t')=\left\{\begin{array}{ll}
		  \psi^f(t)V          \psi^b(t')^{\dagger}\gamma_5 
                          &\quad t<t' \\
		  \psi^b(t)V^{\dagger}\psi^f(t')^{\dagger}\gamma_5 
                          &\quad t>t' \, .
		  \end{array} \right.
\end{eqnarray} 
Here $V$ is a two by two matrix acting on the index that labels
independent solutions.
Inspection of the propagator equation at $t'=t\pm1$ yields in addition
\begin{equation}
{\cal S}(t,t) =          \psi^f_-(t)V          \psi^b(t)^{\dagger}\gamma_5+
                  \psi^b_+(t)V^{\dagger}\psi^f(t)^{\dagger}\gamma_5
\end{equation}
and
\begin{eqnarray} \label{NONTPROP1}
   \psi^f_-(t)V          \psi^b_-(t)^{\dagger}\gamma_5
&=&\psi^b_-(t)V^{\dagger}\psi^f_-(t)^{\dagger}\gamma_5 \, , \\
\label{NONTPROP2}
   \psi^b_+(t)V^{\dagger}\psi^f_+(t)^{\dagger}\gamma_5
&=&\psi^f_+(t)V          \psi^b_+(t)^{\dagger}\gamma_5 \, .
\end{eqnarray}	

To determine $V$ we exploit (\ref{propeq}) at $t=t'$,
\begin{eqnarray} \label{VCOND}
1&=&
  -\psi^f_+(t-1)V          \psi^b_-(t)^{\dagger}\gamma_5
  +\psi^b_+(t-1)V^{\dagger}\psi^f_-(t)^{\dagger}\gamma_5\\\nonumber
&&+\psi^f_-(t+1)V          \psi^b_+(t)^{\dagger}\gamma_5
  -\psi^b_-(t+1)V^{\dagger}\psi^f_+(t)^{\dagger}\gamma_5 \, .
\end{eqnarray}
Now $V$ can be isolated with the help of (\ref{NONTPROP1}), 
(\ref{NONTPROP2}) and
\begin{eqnarray}
   \psi^f  (t-1)^{\dagger} \gamma_5 \psi^f_-(t)
 - \psi^f  (t  )^{\dagger}\gamma_5 \psi^f_+(t-1) = 0 \, ,
\end{eqnarray}	
whose left hand side is shown to be $t$-independent by the 
homogeneous Dirac equation and vanishes for $t=1$.
The final result can be written as
\begin{eqnarray} \label{VINVF}
\left(V^{\dagger}\right)^{-1}&=&
   \psi^f_-(1)^{\dagger}\gamma_5 \psi^b_+(0) .
\end{eqnarray}


\appendix
\renewcommand{\thesection}{C. Tables of expansion coefficients}
\section{}
\renewcommand{\thesection}{C}
In this appendix we list perturbative finite lattice input data
that went into our analysis. Numbers have been truncated such that roundoff
errors affect the last digit only.
\begin{table}[H]
\begin{center}
\begin{tabular}{|c|l|l|l|}\hline
\rule[-2pt]{0cm}{13pt} $I$
    &\multicolumn{1}{|c|}{$p_2^e$}
    &\multicolumn{1}{|c|}{$p_2^f$}
    &\multicolumn{1}{|c|}{$p_2^g$}\\\hline
  3 &  0.01406000332624868 &-0.0182407224873773 &	0.02748117066751495    \\	    
  4 &  0.01838971613803814 &-0.0190839262114088 &	0.02912977487693198    \\	    
  5 &  0.0213244240582013  &-0.0182173759776234 &	0.0272378718296281     \\	    
  6 &  0.0240118520229137  &-0.0171194898351334 &	0.0248632085828385     \\	    
  7 &  0.0266888600910630  &-0.016021619058412  &	0.0226936034816730     \\	    
  8 &  0.029366920333954   &-0.014993754483268  &	0.0208214421618160     \\	    
  9 &  0.032033898489622   &-0.014058323610653  &	0.0192191961300145     \\	    
 10 &  0.034684945753779   &-0.013217310231851  &	0.0178425484742938     \\	    
 11 &  0.037320642901569   &-0.012464162363595  &	0.0166512309905875     \\	    
 12 &  0.039943407218950   &-0.011789492952359  &	0.015612173517774      \\	    
 13 &  0.042555797334290   &-0.011183638952885  &	0.014698916407548      \\	    
 14 &  0.045159985052289   &-0.010637686960534  &	0.013890391073651      \\	    
 15 &  0.04775768717127    &-0.010143794730699  &	0.013169772817157      \\	    
 16 &  0.05035023257788    &-0.009695210528629  &	0.012523543034029      \\	    
 17 &  0.05293864891550    &-0.00928617606722   &	0.011940757139333      \\	    
 18 &  0.05552373638933    &-0.00891179536156   &	0.011412482658586      \\	    
 19 &  0.05810612401759    &-0.00856790398265   &	0.010931369933803      \\	    
 20 &  0.06068631107488    &-0.00825095161082   &	0.010491323683996      \\	    
 21 &  0.06326469747537    &-0.00795790134914   &	0.010087250557429      \\	    
 22 &  0.06584160619894    &-0.00768614535208   &	0.009714863852334      \\	    
 23 &  0.06841730002735    &-0.00743343490502   &	0.009370531360667      \\	    
 24 &  0.07099199418417    &-0.00719782275317   &	0.009051155905983      \\	    
 25 &  0.07356586599137    &-0.00697761558627   &	0.008754080827932      \\	    
 26 &  0.07613906232516    &-0.00677133484944   &	0.008477014637251      \\	    
 27 &  0.0787117066394     &-0.00657768434489   &	0.008217970511475      \\	    
 28 &  0.0812838974813     &-0.00639552336330   &	0.007975217365077      \\	    
 29 &  0.0838557242252     &-0.00622384432090   &	0.007747240012744      \\	    
 30 &  0.0864272578552     &-0.00606175407446   &	0.007532706527406      \\	    
 31 &  0.0889985593460     &-0.00590845824714   &	0.007330441330203      \\	    
 32 &  0.0915696803402     &-0.00576324802706   &	0.007139402877142      \\\hline
\end{tabular}
\end{center}
\caption[]{\label{ptable1}List of $I$-dependent coefficients  
$p_2^e$, $p_2^f$ and $p_2^g$ for $\csw^{(0)}=1$.}
\end{table}
\clearpage
\begin{table}[H]
\begin{center}
\begin{tabular}{|c|l|l|l|}\hline
\rule[-2pt]{0cm}{13pt} $I$
    &\multicolumn{1}{|c|}{$p_2^h$}
    &\multicolumn{1}{|c|}{$p_2^i$}
    &\multicolumn{1}{|c|}{$p_2^j$}\\\hline
  3 &  -0.0472978469844831 &0.02748117066751495 &	 0.00036913190875759   \\	    
  4 &  -0.0522497820522355 &0.0406644614491804  &	-0.00019279262914480   \\	    
  5 &  -0.0544132909735601 &0.0518711757366513  &	 0.000062922280309919  \\	    
  6 &  -0.0555160735660078 &0.0625335624963019  &	 0.00024693874361661   \\	    
  7 &  -0.0561252173383393 &0.0729651117622647  &	 0.00032829652675201   \\	    
  8 &  -0.0564756935876461 &0.0832353491549959  &	 0.00035889080255099   \\	    
  9 &  -0.0566798625246374 &0.0933761748042426  &	 0.00036802392907154   \\	    
 10 &  -0.0567967356276605 &0.103414110425804   &	 0.00036829902719101   \\	    
 11 &  -0.0568594539035442 &0.113371699857364   &	 0.00036493229776592   \\	    
 12 &  -0.056887606070495  &0.123266779814805   &	 0.00036013282273909   \\	    
 13 &  -0.056893249514880  &0.133112866383861   &	 0.00035486899918541   \\	    
 14 &  -0.056884053136243  &0.142920056704081   &	 0.00034957573793481   \\	    
 15 &  -0.056865026658050  &0.152695918495843   &	 0.00034444805110225   \\	    
 16 &  -0.056839514713774  &0.162446191871116   &	 0.00033956882844280   \\	    
 17 &  -0.056809789874931  &0.172175297258459   &	 0.0003349670866427    \\	    
 18 &  -0.056777417868152  &0.181886691902227   &	 0.0003306456480143    \\	    
 19 &  -0.056743488785825  &0.191583119155590   &	 0.0003265947619586    \\	    
 20 &  -0.056708767026702  &0.201266784482244   &	 0.0003227989929269    \\	    
 21 &  -0.056673790607995  &0.210939481638856   &	 0.0003192407628410    \\	    
 22 &  -0.056638938178912  &0.220602684697517   &	 0.0003159021777408    \\	    
 23 &  -0.056604474988832  &0.230257616284550   &	 0.0003127659514888    \\	    
 24 &  -0.056570584880899  &0.239905298968026   &	 0.0003098158454325    \\	    
 25 &  -0.056537392847533  &0.249546594496119   &	 0.0003070368462269    \\	    
 26 &  -0.056504981113718  &0.259182234133082   &	 0.0003044152025451    \\	    
 27 &  -0.056473400720539  &0.268812842375392   &	 0.0003019383874879    \\	    
 28 &  -0.056442679941137  &0.278438955679912   &	 0.0002995950240854    \\	    
 29 &  -0.056412830441785  &0.288061037388928   &	 0.0002973747949003    \\	    
 30 &  -0.056383851821358  &0.297679489724326   &	 0.0002952683474247    \\	    
 31 &  -0.056355734973936  &0.307294663501017   &	 0.0002932672016335    \\	    
 32 &  -0.056328464590232  &0.31690686604973    &	 0.0002913636629602    \\\hline  
\end{tabular}
\end{center}
\caption[]{\label{ptable2}List of $I$-dependent coefficients  
$p_2^h$, $p_2^i$ and $p_2^j$ for $\csw^{(0)}=1$.}
\end{table}
\clearpage
\begin{table}[H]
\begin{center}
\begin{tabular}{|c|l|l|l|}\hline
\rule[-2pt]{0cm}{13pt} $I$
    &\multicolumn{1}{|c|}{$p_2^e$}
    &\multicolumn{1}{|c|}{$p_2^i$}
    &\multicolumn{1}{|c|}{$p_2^j$}\\\hline
  3 &  0.002930793635704460& 0.01113927564510310& -0.00081995809255674 \\
  4 &  0.00648721649586021 &	0.02176506549404950&	-0.00095088311404386 \\	    
  5 &  0.01042316816041390 &	0.0326987791899857 &	-0.00047704024765898 \\	    
  6 &  0.0147884221405374  &	0.0441250945952050 &	-0.00017002734632304 \\	    
  7 &  0.0194100790766652  &	0.0557968441900837 &	-0.000012571985852322\\ 	  
  8 &  0.0241548411409164  &	0.0675196648718815 &	 0.000069240759237721\\ 	  
  9 &  0.028948291534668   &	0.0791944332878618 &	 0.00011551757360719 \\	    
 10 &  0.033752733950337   &	0.0907802022873036 &	 0.00014436115981723 \\	    
 11 &  0.038549666950648   &	0.102264260790802  &	 0.00016392305375706 \\	    
 12 &  0.043330159467278   &	0.113646262219098  &	 0.00017808308496443 \\	    
 13 &  0.048090083743491   &	0.124930880299287  &	 0.00018883175180516 \\	    
 14 &  0.05282779555547    &	0.136124585602541  &	 0.00019727388656343 \\	    
 15 &  0.05754297570840    &	0.147234255392045  &	 0.00020407010116580 \\	    
 16 &  0.06223602393897    &	0.158266580822338  &	 0.00020964219355829 \\	    
 17 &  0.06690772585583    &	0.169227824652203  &	 0.00021427470438615 \\	    
 18 &  0.07155906252343    &	0.180123737178240  &	 0.00021816831762828 \\	    
 19 &  0.07619109889499    &	0.190959545294390  &	 0.00022146967768869 \\	    
 20 &  0.08080491792206    &	0.201739975291261  &	 0.00022428900150278 \\	    
 21 &  0.08540158202310    &	0.212469290287585  &	 0.00022671102155128 \\	    
 22 &  0.08998211127935    &	0.223151332698423  &	 0.00022880209115049 \\	    
 23 &  0.09454747195837    &	0.233789566873300  &	 0.00023061497375955 \\	    
 24 &  0.0990985714169     &	0.244387119503159  &	 0.00023219217419543 \\	    
 25 &  0.1036362569094     &	0.254946816719166  &	 0.0002335683174515  \\	    
 26 &  0.1081613167382     &	0.265471217523167  &	 0.0002347718854825  \\	    
 27 &  0.1126744827534     &	0.275962643571370  &	 0.0002358265093730  \\	    
 28 &  0.1171764335743     &	0.286423205525826  &	 0.0002367519464219  \\	    
 29 &  0.1216677981308     &	0.296854826274766  &	 0.0002375648294884  \\	    
 30 &  0.1261491592962     &	0.307259261350171  &	 0.0002382792488587  \\	    
 31 &  0.1306210574330     &	0.31763811686605   &	 0.0002389072090909  \\	    
 32 &  0.1350839937928     &	0.32799286527967   &	 0.0002394589912651  \\\hline
\end{tabular}
\end{center}
\caption[]{\label{ptable3}List of $I$-dependent coefficients  
$p_2^e$, $p_2^i$ and $p_2^j$ for $\csw^{(0)}=0$.}
\end{table}
\clearpage


\appendix
\renewcommand{\thesection}{D. Extrapolation of lattice perturbation
theory}
\section{}
\renewcommand{\thesection}{D}

In this appendix we discuss a modified method to extract the
continuum limit behavior of lattice Feynman diagrams. 
That is we assume to have precise results $F(I)$ 
(for a given diagram or a sum of diagrams) 
for a range of $I=L/a$, $I_1<I_2<\dots <I_n$
(typical subsets of $I$ are $2,\dots,32$), and the goal is to
determine the leading and subleading behavior as $I\to\infty$.

The precision of the data is limited by roundoff effects. We treat
the roundoff errors $\delta_F(I)$ as normally distributed superimposed
noise, independent for different $I$. The most optimistic assumption
for results from double precision (real*8) arithmetic would be
\be
\delta_F(I)=\epsilon |F(I)|\,,\,\,\,\,\,\,\epsilon\sim 10^{-14}\,,
\ee
but in general this is an underestimation since the evaluation of
the diagrams involve large sums of terms of different signs.
A comparison of double versus single (or quadruple) precision
for a range of $I$ can be used to derive a more realistic
$I-$ dependent $\epsilon$. (In our particular case we estimated
a growth $\propto I^3$ for most contributions). 

For the extrapolation we assume an asymptotic expansion in 
functions $\Bigl\{f_k(I)\,,k=1,2,\dots,n_f\Bigr\}$ where 
successive terms become ``smaller" with increasing index $k$,
\be
F(I)=\sum_{k=1}^{n_f}\alpha_k f_k(I)+R(I)\,,
\ee 
where $|R(I)/f_{n_f}|\to 0$ as $I\to\infty$. Of course, we have
in mind something like $\Bigl\{f_1(I),f_2(I),\dots\Bigr\}
=1,\ln I/I,\dots ,\ln^{\mu} I/I^{\nu},\dots\Bigr\}$. Denoting the
$n$ data values for $F(I)$ as an $n$-dimensional column vector,
the above equation reads
\be
F=f\alpha+R\,,
\ee
where $f$ (evaluated for the $n$ values of $I$ for which we have data)
is regarded as an $n\times n_f$ matrix and $\alpha$ is an
$n_f$-dimensional column vector that we want to determine.

In ref.~\cite{LWblock} a recursive blocking technique was proposed
to determine $\alpha$ which was claimed superior to making
least square fits. In the following we shall find that blocking 
can actually be considered as a particular way of fitting with
generalizations however being potentially more convenient.

We determine $\alpha$ by minimizing a quadratic form in the residues
\be
\chi^2=(F-f\alpha)^{\rm T}W^2(F-f\alpha)\,.
\label{chisq}
\ee
Here $W^2$ is an $n$-dimensional matrix of positive weights.
It can be used to emphasize residues at small or large $I$
in the minimization. Small $I$ are less affected by roundoff,
leading to less roundoff errors induced in $\alpha$. On the
other hand for larger $I$ the asymptotic expansion holds
to a better degree, and hence here smaller systematic errors
are expected.  
In fact for our data we found little advantage in 
setting $W_{IJ}=\delta_{IJ}I^z,-3\le z\le 3$; and thus we
finally used a flat weight $W=1$ and just changed the $I$ range analyzed
to observe convergence by moving $I_{\rm min}$ while always
keeping $I_{\rm max}$ at the highest available size. 
The freedom in choice of $W$ could however be useful in different
situations and hence we keep it general in the following discussion.

Minimization of (\ref{chisq}) leads to the equation
\be
f^{\rm T}W^2f\alpha=f^{\rm T}W^2F\,.
\ee
We assume that the columns of $Wf$ are linearly independent thus
spanning an $n_f$-dimensional subspace ($n_f<n$ must always hold).
Let $P$ be the projector on to this subspace. Then
\be
Wf\alpha=PWF\,,
\label{proj}
\ee
is an equivalent equation that fixes $\alpha$. A very stable and
convenient way to solve for $\alpha$ is to construct the
{\it singular value decomposition} for $Wf$ \cite{numrec}.
In a very simple version it amounts to a factorization
\be
Wf=USV^{\rm T}\,,
\ee
where $U$ is a column-orthonormal $n\times n_f$ matrix obeying
\be
U^{\rm T}U=1\,;\,\,\,\,UU^{\rm T}=P\,,
\ee
$S$ is diagonal and $V$ orthonormal, both of size $n_f\times n_f$.
Then we get
\be
\alpha=VS^{-1}U^{\rm T}WF\,.
\label{soln}
\ee
By simple error propagation the roundoff error of $\alpha$ is
\be
\delta^2_{\alpha_k}=\sum_I \Bigl( VS^{-1}
U^{\rm T})^2_{kI}\delta^2_{F(I)}\,.
\ee

A simple property of (\ref{proj}) is the following. Imagine
changing $F$ by a component proportional to one of the functions
$f_k$ included in the fit. Then it is easy to see that only the
corresponding $\alpha_k$ changes. In other words an estimate
of $\alpha_1$ from (\ref{soln}) is only uncertain due to 
components beyond $f_1,\dots,f_{n_f}$ contained in $F$.
Then it is clear that if we choose the $I$-window minimal 
($n=n_f$ successive values) the solution is equivalent
to a blocking procedure cancelling $n_f-1$ components to
isolate say $\alpha_1$.

The estimation of the systematic errors is the most delicate 
problem in our context. After considerable experimentation
we propose the following procedure. We assume that the
remainder $R$ can be modelled by a linear combination
of the $n_r$ functions $f_{n_f+1},f_{n_f+2},\dots,f_{n_f+n_r}$ 
following those included in the fit. Typically in our case
these would be the $n_r=3$ functions $\ln^2 I/I^m,\ln I/I^m,1/I^m$,
if O($1/I^{m-1}$) is the last order in the fit. If they could
be fitted individually (that is distinguished), they would be
included in the analysis. Instead, for our error analysis we 
make $n_r$ separate fits including {\it only one of them}
per fit in addition to the $f_i, i=1,\dots,n_f$. In this way we 
get the coefficients $A_1,\dots,A_{n_r}$ of the extra term
in each case. The model remainders
$A_1f_{n_f+1},A_2f_{n_f+2},\dots,A_{n_r}f_{n_f+n_r}$ are fitted
in the same way as the data (\ref{soln}), and the maximal
fit-coefficient of the $n_r$ cases is taken as the systematic error
$d_{\alpha_k}$ of $\alpha_k$.

We found that these errors based on the leading {\it unincluded}
contribution are smaller in most (but not all) cases compared to the
previous method focussing on the last {\it included} term.
The present errors seem more realistic in the sense that
expected relations (like coefficients of the $\beta$-function)
hold to an accuracy still conservatively but not grossly
more accurate than the estimated errors. 

\vfill
\eject

\end{document}